\documentclass[
 reprint,
 amsmath,amssymb,
 aps,
pra,
floatfix,
]{revtex4-2}

\usepackage{graphicx}
\usepackage{dcolumn}
\usepackage{amsmath}
\usepackage{amsthm}
\usepackage{enumitem}
\usepackage{bm}
\usepackage{hyperref}
\usepackage{physics}
\usepackage{xcolor}
\usepackage{graphicx}
\usepackage{lipsum}
\usepackage{tikz}
\usetikzlibrary{quantikz2}
\usepackage{listings}
\usepackage[caption=false]{subfig}
\usepackage[utf8]{inputenc}
\usepackage[normalem]{ulem}
\hypersetup{colorlinks,citecolor=blue,linkcolor=blue,urlcolor=blue}

\newcommand{\cC}{\mathcal{C}}
\newcommand{\cD}{\mathcal{D}}
\newcommand{\cE}{\mathcal{E}}

\newcommand{\cN}{\mathcal{N}}

\newcommand{\cS}{\mathcal{S}}
\newcommand{\cT}{\mathcal{T}}

\newcommand{\cZ}{\mathcal{Z}}

\begin{document}

\preprint{APS/123-QED}

\title{Benchmarking logical three-qubit quantum Fourier transform encoded in the Steane code on a trapped-ion quantum computer}

\author{Karl Mayer}
\author{Ciar\'{a}n Ryan-Anderson}
\author{Natalie Brown}
\author{Elijah Durso-Sabina}
\author{Charles H. Baldwin}
\author{David Hayes}
\author{Joan M. Dreiling}
\author{Cameron Foltz}
\author{John P. Gaebler}
\author{Thomas M. Gatterman}
\author{Justin A. Gerber}
\author{Kevin Gilmore}
\author{Dan Gresh}
\author{Nathan Hewitt}
\author{Chandler V. Horst}
\author{Jacob Johansen}
\author{Tanner Mengle}
\author{Michael Mills}
\author{Steven A. Moses}
\author{Peter E. Siegfried}
\author{Brian Neyenhuis}
\author{Juan Pino}
\author{Russell Stutz}

\affiliation{Quantinuum}
\date{\today}

\begin{abstract}
We implement logically encoded three-qubit circuits for the quantum Fourier transform (QFT), using the [[7,1,3]] Steane code,
and benchmark the circuits on the Quantinuum H2-1 trapped-ion quantum computer.
The circuits require multiple logical two-qubit gates,
which are implemented transversally,
as well as logical non-Clifford single-qubit rotations,
which are performed by non-fault-tolerant state preparation followed by a teleportation gadget.
First, we benchmark individual logical components using randomized benchmarking for the logical two-qubit gate, and a Ramsey-type experiment for the logical $T$ gate.
We then implement the full QFT circuit,
using two different methods for performing a logical control-$T$,
and benchmark the circuits by applying it to each basis state in a set of bases that is sufficient to lower bound the process fidelity.
We compare the logical QFT benchmark results to predictions based on the logical component benchmarks.

\end{abstract}

\maketitle

\section{Introduction}\label{sec: 1}

It is generally believed that quantum
error correction (QEC) will be required for future large-scale quantum computation.
This involves encoding logical qubits into
a higher dimensional space comprising a larger number of physical qubits,
applying gates at the logical level,
and performing rounds of syndrome extraction and error correction.
Additionally, fault-tolerant circuit design will be needed
to ensure that errors do not spread uncontrollably
during an arbitrary computation.
An important criterion for useful QEC is that
for every circuit in a family of interest,
the error rate for an encoded version of that circuit must be lower than that of an unencoded version run on the same hardware~\cite{Gottesman2016}.
Recent progress towards that goal,
enabled by advances in scalability and improved gate fidelity,
has allowed for direct comparison between small encoded circuits and unencoded circuits~\cite{ryananderson2021, ryananderson2022, self2022, yamamoto2024, Wang2023, gupta2024, Bluvstein2024, dasilva2024}.

In currently available devices,
it is not \textit{a priori} clear which codes and protocols (initialization, gates, syndrome extraction, etc.) lead to the best circuit performance,
and in order to weigh the different options available we must rely on quantum benchmarking tools.
At the physical level, component benchmarking has been used to characterize and quantify the performance of individual parts of a quantum circuit,
whereas system-level benchmarks assess the overall 
ability to execute quantum circuits with high fidelity.
Assessing whether the error rates of individually benchmarked components account for the error rate of the full circuit is also important so that one can be confident in results when quantum processors perform novel calculations.

Here, we report on the experimental extension of these methods to the logical level. We use the Quantinuum H-series quantum processors~\cite{Pino2020, Moses2023} to experimentally investigate both component-level (individual gates) and system-level (complex circuits) benchmarks of the [[7,1,3]] Steane code~\cite{Steane1996}. At the component-level, we benchmark the most error-prone logical gates,
namely the transversal two-qubit logical CNOT,
and the non-Clifford logical $T$ gate.
We benchmark these components by two-qubit randomized benchmarking (RB)~\cite{Knill2008, Magesan2011},
and a protocol from Ref.~\cite{Piveteau2021} for estimating
the fidelity of a logical $T$ performed by a teleportation gadget.
At the system-level, we perform an encoded three-qubit quantum Fourier transform (QFT), chosen because of its importance as a subroutine and because it requires
all the logical primitives for universal
quantum computation.
With 7 physical qubits per logical qubit block,
plus another 7 for a logical ancilla qubit needed
to prepare non-Clifford magic states,
the three-qubit QFT is the largest that fits within the 
current 32-qubit H2-1 architecture~\cite{Moses2023}.
All of these experiments were designed using a software framework developed in-house that we call Simple Logical Representation (SLR)~\cite{PECOS}.
We stress that we do not employ fault-tolerant non-Clifford gates or rounds of QEC,
though it is worth noting that real-time QEC is done based on destructive measurements used in the gate teleportation gadgets. 
In this work, the focus is on benchmarking the logical encoding and
assessing whether component-level errors account
for the full system performance at the logical level.  

We measure logical two-qubit gate average fidelities 
0.9991(2) on H1-1 and 0.9980(8) on H2-1,
which is at or near the break-even level compared to physical two-qubit gates.
For the logical $T$ gate on H2-1, we obtain a much
lower average fidelity of 0.990(1).
We implement the logical QFT in two different ways,
based on two methods for encoding a logical control-$T$.
For both methods, we apply the QFT to all basis states
in both the computational basis,
and a rotated Fourier basis, from which a lower bound
on the logical process fidelity can be computed.
For the higher performing of these two methods,
we obtain average output state fidelities of 0.78(1) and 0.66(1) in these two bases.
When allowing post-selection on syndrome information obtained from the teleportation gadgets,
the average output state fidelities increase to 0.89(1) and 0.77(2),
respectively.
These numbers are lower than what we would 
obtain from an unencoded QFT circuit, with the largest error stemming from the $T$ gates.
We conclude that either higher distance codes or fully fault-tolerant
circuit construction (or both) would be required for the logical circuit to outperform a physical circuit on our present hardware.
Finally, we model the logical QFT circuits
using the logical component-level benchmarking results.
We find that the logical gate errors explain part but not all of the circuit error,
and closing this gap will be crucial in developing
large-scale devices capable of running algorithms at the logical level.

The outline of the paper is as follows.
Sec.~\ref{sec: 2} provides background
information on the [[7,1,3]] code
and details the logical primitives used in this work.
We assume familiarity with the
stabilizer formalism of quantum error correction~\cite{gottesman1997stabilizer}.
In Sec.~\ref{sec: 3} we describe the experimental logical component benchmarking of the transversal two-qubit logical CNOT gate,
and the non-Clifford logical $T$ gate.
Section~\ref{sec: 4} contains the details of the logical QFT implementations,
the procedure for obtaining process fidelity lower bounds,
and the experimental results.
We conclude in Sec.~\ref{sec: 5} with a discussion and outlook.

\section{Quantum error correction primitives}\label{sec: 2}

\subsection{The 7-Qubit Steane Code}

In this work we use the [[7,1,3]] color code, also known as the Steane code~\cite{Steane1996}.
This code uses 7 physical qubits to encode one logical qubit to distance 3,
meaning that it is capable of detecting and correcting all single-qubit errors.
The Steane code is an example of a color code,
which is a class of topological codes defined by a 3-coloring of a planar graph,
where each vertex of the graph corresponds to a qubit,
and there is an $X$-type and $Z$-type stabilizer for each plaquette~\cite{bombin2007topological, Landahl2011}.
In our convention a generating set of stabilizers and logical operators $\overline{X}$ and $\overline{Z}$ are given by
\begin{align}\label{eq: def color code}
    s_1&=XXXXIII,\notag\\
    s_2&=IXXIXXI,\notag\\
    s_3&=IIXXIXX,\notag\\
    s_4&=ZZZZIII,\notag\\
    s_5&=IZZIZZI,\\
    s_6&=IIZZIZZ, \notag\\
    \notag \\
    \overline{X} & = IIIIXXX, \notag \\
    \overline{Z} & = IIIIZZZ. \notag
\end{align}
Here, we use an overhead bar when 
needed to distinguish a logical operator from a physical operator. 
The color code is a strong CSS code, meaning that the $X$-type and $Z$-type stabilizers have identical support~\cite{landahl2014quantum}.
Consequentially, all logical Clifford operations are
transversal.
To see this, consider the action of  
a transversal $\overline{H}$ and $\overline{S}$ on stabilizers and logical operators.
Applying $\overline{H}=H^{\otimes7}$ to the stabilizers listed in Eq.~\eqref{eq: def color code},
we see that $\overline{H}$ preserves the stabilizer group,
and is therefore a logical operator.
For example, $s_1$ transforms into $s_4$ and vice versa. 
Furthermore, when we conjugate $\overline{X}$ in Eq.~\eqref{eq: def color code} by $\overline{H}$, we get $\overline{Z}$, and vice versa, so the logical Pauli operators also transform appropriately. 
The transversal $\overline{S}$ is implemented by applying $S^\dag$ to all physical qubits.
Finally, a logical two-qubit Clifford, such as $\overline {\mathrm{CNOT}}$, 
is given by a physical CNOT between each pair of corresponding qubits between two code blocks.
Since $H$, $S$, and CNOT generate the Clifford group,
it follows that all logical Clifford operators are transversal.
For more details see ~\cite{ryananderson2021}.

Throughout this paper we use the term ``fault-tolerance," which carries various meanings, depending on the context. Here, we describe circuit design principles as being fault-tolerant if they control the spread of ``faults" in a QEC circuit.
Precisely, we call a sub-circuit fault-tolerant if
no error on a single qubit in a code block can propagate to an error that is uncorrectable
by a round of ideal syndrome extraction and correction~\cite{gottesman2009}.
An example of an uncorrectable error is one that
is equivalent to
a non-trivial logical operator, that is, an element of $\cN\setminus\cS$,
where $\cS$ is the stabilizer group generated by Eq.~\eqref{eq: def color code},
and $\cN=\{g:gsg^{\dag}\in\cS\ \forall s\in\cS\}$
is the normalizer of $\cS$.
Such errors are undetectable and hence uncorrectable.
The transversality of logical Clifford operations in the color code guarantees their fault-tolerance.

\begin{figure}[h]
    \centering
    \includegraphics[width=0.3\textwidth]{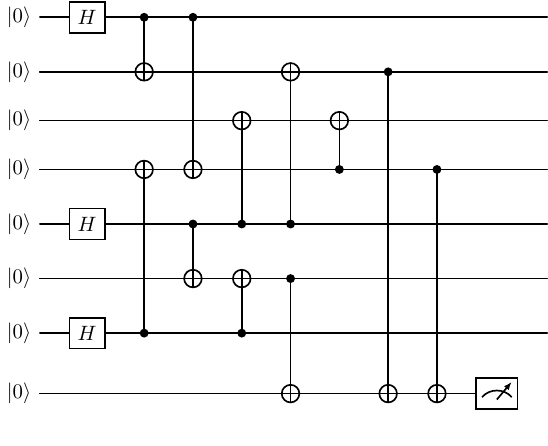}
    \caption{Fault-tolerant repeat-until-success $\ket{\overline{0}}$ initialization circuit.
If the ancilla qubit measurement result is `0',
then the initialization succeeds.
Otherwise, the circuit is repeated, up to a limit of three repetitions.}
    \label{fig: FT encoding circuit}
\end{figure}

Fault-tolerant logical qubit initialization in the Steane code is performed by an encoding circuit shown in Fig.~\ref{fig: FT encoding circuit}, which uses a logical measurement scheme to achieve fault-tolerance~\cite{goto2016minimizing}.
The encoding circuit itself is not fault-tolerant, 
since the occurrence of a single-qubit error in certain locations produces a two-qubit error. 
To ensure fault-tolerance, the circuit uses an ancilla physical qubit to make a $\overline{Z}$ measurement that 
detects any two-qubit error that might occur.  
A measurement result of `0' indicates successful preparation of
the logical state $\ket{\overline{0}}$, otherwise the circuit is repeated.
In our experiments,
we set a limit of three repetitions of this repeat-until-success protocol
and post-select to discard shots that never succeeded.
The discard rate from this initialization is small;
at most a few percent of the shots in our
benchmarking experiments are discarded. 
The choice of the limit three is motivated from our earlier work~\cite{ryananderson2021}.

Logical qubit measurement in a Pauli basis is performed by destructively measuring each physical qubit in that basis.
From this measurement, one obtains the parities of both the logical operator,
and the corresponding stabilizer generators,
by calculating the parity of the qubits that support each operator.
The stabilizer information then gives a syndrome which can be decoded, and a final correction can be applied to the raw logical outcome.

\begin{figure}[h]
    \centering
    \subfloat[\label{fig: method 1}]{
        \centering
        \includegraphics[width=0.22\textwidth]{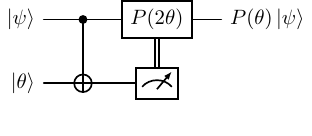}}
    \hfill
    \subfloat[\label{fig: method 2}]{
        \centering
        \includegraphics[width=0.22\textwidth]{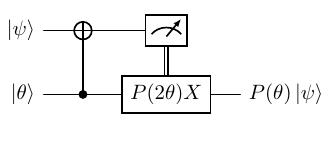}}
    \caption{Two versions of a teleportation gadget
    for performing a logical phase gate,
    which we refer to as (a) method one and (b) method two.
The circuits use an ancilla prepared in $\ket{\theta}=\frac{1}{\sqrt{2}}\big(\ket{0}+e^{i\theta}\ket{1}\big)$,
and output the state $P(\theta)\ket{\psi}$,
where $P(\theta)=\mathrm{diag}(1,e^{i\theta})$.
Taking $\theta=\pi/4$ yields the $T$ gate.
}
    \label{fig: teleporation gadget2}
\end{figure}

\begin{figure}[h]
    \centering
    \includegraphics[width=0.3\textwidth]{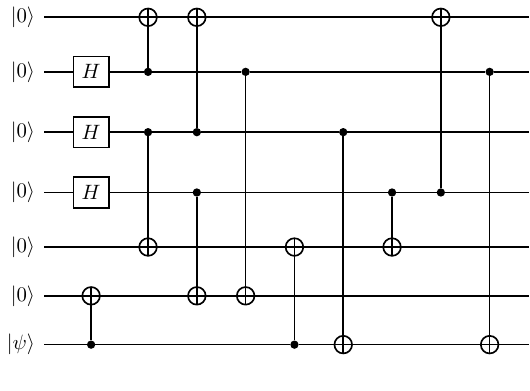}
    \caption{Non-fault-tolerant logical encoding circuit for the Steane code.
The final qubit in the quantum register is prepared in $\ket{\psi}$ and the resulting logical state is $\ket{\overline{\psi}}$.}
    \label{fig: encoding circ}
\end{figure}

The Eastin-Knill theorem shows that a universal transversal logical gate set does not exist~\cite{Eastin2009}.
To complete a universal logical gate set,
we perform the gate $\overline T$ by preparing the logical state
$\ket{\overline T}=\frac{1}{\sqrt{2}}\big(\ket{\overline{0}}+e^{i\pi/4}\ket{\overline{1}}\big)$
and applying either of the teleportation gadgets shown in Fig.~\ref{fig: teleporation gadget2}.
More generally, we can apply a logical $Z$-rotation
given by $P(\theta)=\mathrm{diag}(1,e^{i\theta})$,
by preparing $\ket{\overline\theta}=\frac{1}{\sqrt{2}}\big(\ket{\overline{0}}+e^{i\theta}\ket{\overline{1}}\big)$ non-fault-tolerantly,
followed by a teleporation gadget.
The circuit for preparing $\ket{\overline\theta}$ non-fault-tolerantly is shown in Fig.~\ref{fig: encoding circ}.
The two gadgets in Fig.~\ref{fig: teleporation gadget2} are logically equivalent and differ
only in whether the initial logical data or
ancilla qubit block ends up as the output data qubit block.
We benchmark both gadgets as described in the next section.

The teleportation gadgets use a logical measurement
and gate conditioned on the logical outcome.
In our implementation, we use the syndrome information obtained from the measurement to obtain a decoded logical outcome in real-time,
and condition the gate on the decoded logical outcome.
It is also possible to discard shots in which the syndromes from this
measurement indicate an error,
essentially using the code for quantum error detection (QED).
However, this post-selection is not scalable as the
measurement is made after the logical ancilla qubit has been entangled with the logical data qubit,
and so a detection event requires discarding the entire computation.
The shot overhead will therefore scale exponentially in the logical $T$ count.
Since the logical QFT circuits in this work are small enough,
we compare the results with and without post-selection on the teleportation gadget measurements.

The $\ket{\overline T}$ state preparation circuit can be made
fault-tolerant by using an ancilla qubit to measure the projector $\ket{\overline{T}}\bra{\overline{T}}$
with an additional flag qubit to detect
errors in the measurement that spread to uncorrectable errors,
followed by a round of QEC~\cite{Chamberland2019, Postler2022}.
This scheme can be applied in a repeat-until-success (RUS) protocol,
combined with post-selection to discard shots
in which a limit on the number of unsuccessful attempts was reached.
In this work, however, we only implement the non-fault-tolerant version of $\overline{T}$ on quantum hardware.
This is because simulations (detailed in Appendix~\ref{appendix B}) using the device emulator suggest that
a partially-fault-tolerant $\overline{T}$
(that includes the flagged measurement but not the QEC round)
would have a lower fidelity than the non-fault-tolerant $\overline{T}$,
due to a combination of physical two-qubit gate error and
memory error incurred during the flag measurement.
We note that a time-efficient protocol could use a logical $T$-state factory where many attempts to prepare $\ket{\overline{T}}$ are done in parallel.
However, this would require more qubit resources than are currently available in the H-series computers.
We leave an experimental verification of the simulations and a detailed study of full fault-tolerance to future work.

\subsection{Simple Logical Representation}

To ease the construction of logically encoded circuits,
we developed a software framework that we call Simple Logical Representation (SLR)~\cite{PECOS}.
In SLR, a logical circuit representation can be
built in a Pythonic environment,
similar to other circuit programming frameworks such as pytket~\cite{Sivarajah2021} or Qiskit~\cite{Qiskit}.
An SLR object draws from a custom
in-house library of QEC primitives to output a program built from OpenQasm 2.0~\cite{Cross2017} and the Quantinuum OpenQasm 2.0 extensions~\cite{ryananderson2021},
building off of prior experimental work that demonstrated real-time quantum error correction~\cite{ryananderson2021}
and a logically encoded transversal CNOT gate~\cite{ryananderson2022}.

As an example of SLR code, a $\overline{T}$ gate
as implemented by the teleportation gadget of
Fig.~\ref{fig: method 1} is appended to
an SLR program as follows:
\begin{lstlisting}[language=python, basicstyle=\small]

program.extend(

    InitTPlusNonFT(a),
    CX(q,a),
    MeasDecode(a,c,cLogRaw,cLog,cSyn),
    If(cLog==1).Then(S(q)),
)

\end{lstlisting}
Here `q' and `a'
are quantum registers for the logical data and ancilla qubits, respectively.
The object \lstinline{InitTPlusNonFT} is used to initialize the ancilla in the state $\ket{\overline{T}}$
via the non-fault-tolerant encoding circuit in Fig.~\ref{fig: encoding circ}.
A classical register `c' is used for the measurement of the qubits in register `a', and 
`cLog' is a single classical bit used to record the decoded value of $\overline{Z}$ from
the raw logical output `cLogRaw' and the syndromes `cSyn'.
In this example, the objects \lstinline{InitTPlusNonFT},  \lstinline{CX}, and \lstinline{MeasDecode} are all imported from the Steane code QEC library,
whereas \lstinline{If} and \lstinline{Then} are SLR objects
that can be used with any QEC code.
A more complete SLR program is given in Appendix~\ref{appendix C}.
From the above example the benefits of SLR are clear:
one can program circuits at the logical level once the various logical
primitives have been defined.
An SLR program has a \lstinline{.qasm()} method that returns a string of OpenQasm 2.0 code,
which can be submitted to the Quantinuum H-series hardware via a submission api.

\section{Logical component benchmarking}\label{sec: 3}

In this section, we describe the protocols for benchmarking
logical gates, and present the results from the Quantinuum H-series hardware.
The two protocols covered in this section are logical two-qubit randomized benchmarking (RB), and logical $T$ gate benchmarking.
We focus on these two logical operations since
they are composed of physical two-qubit gates, and therefore are a leading source of circuit error.
We run the experiments on both the devices H1-1 and H2-1.
The results of our logical benchmarking experiments are shown
in Table.~\ref{tab: benchmarks}.
The figure of merit we use for logical gate
benchmarking is the average fidelity,
defined as
\begin{equation}
    F_{avg}(\cE,U)=\int d\psi\bra{\psi}U^{\dag}\cE\big(\ket{\psi}\bra{\psi}\big)U\ket{\psi},
\end{equation}
where $U$ is a target unitary, $\cE$ is the actual noisy quantum process that is implemented,
and the integral is with respect to the measure on
pure states induced by the Haar unitary measure~\cite{Mele2023}.
In the case where $U$ is a logical operator encoded in a QEC code, the integral is taken over the code space,
which is a subspace of the physical qubit Hilbert space.

\begin{table}
\begin{ruledtabular}
\begin{tabular}{lcc}
Experiment & Machine & Average Fidelity \\ \hline
Two-qubit RB          &    H1-1    &   0.9991(2)    \\
Two-qubit RB          &    H2-1    &   0.9980(8)    \\
$T$ method one        & H2-1 & 0.981(1) \\
$T$ method two        & H2-1 & 0.990(1) \\
$T$ method one (P.S.) & H2-1 & 0.985(3) \\
$T$ method two (P.S.) & H2-1 & 0.995(1) \\

\end{tabular}
\end{ruledtabular}
\caption{Logical component benchmarking results. For two-qubit RB, the fidelity listed in the table is the estimated average fidelity of the transversal logical CNOT.
For the logical $T$ experiments,
methods one and two refer to the two methods for
performing a teleportation gadget shown in Fig.~\ref{fig: teleporation gadget2},
and P.S. refers to post-selecting the results
on the syndromes obtained from the logical measurement in the teleportation gadgets.
}
\label{tab: benchmarks}
\end{table}

\subsection{Randomized Benchmarking of Logical CNOT}

We first benchmark the logical transversal CNOT gate,
via two-qubit randomized benchmarking (RB)~\cite{Magesan2011}.
Logical RB was previously considered in Ref.~\cite{Combes2017}.
We generate the full two-qubit Clifford group $\cC_2$ using
a generating set of single-qubit Paulis, $H$, $S$ and $S^{\dag}$, as well as CNOT.
The group $\cC_2$ has 11,520 elements. The RB circuits are
constructed by applying a sequence $(C_i)_{i=1}^L$ of Cliffords, each drawn uniformly at random, followed by $C_{inv}=C_1^{-1}\dots C_L^{-1}$,
which is a Clifford group element equal to the inverse of the sequence.
The circuits use a total of 16 physical qubits: 7 for each logical qubit block and
one ancilla for each logical qubit for fault-tolerant initialization.

\begin{figure}[h]
\includegraphics[width=0.45\textwidth]{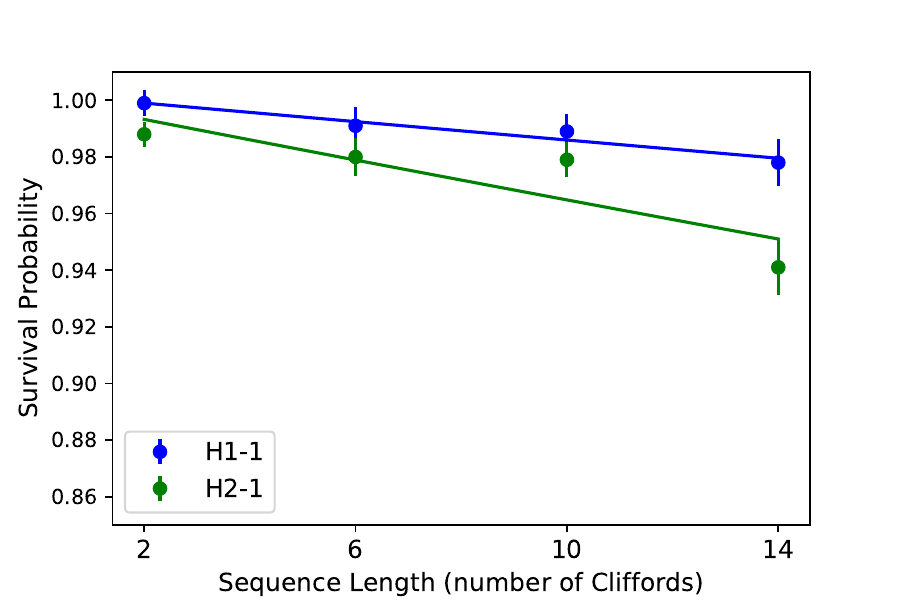}
\caption{Logical two-qubit RB decay curves for the devices H1-1 and H2-1.
Ten random circuits per sequence length
were chosen,
and all circuits were run with 100 shots and submitted in a random order.
The curves are fit to Eq.~\eqref{eq: RB decay},
and the estimated average fidelities per logical CNOT are listed in Table~\ref{tab: benchmarks}.
}
\label{TQRB}
\end{figure}

We use sequence lengths $L\in\{2, 6, 10, 14\}$ and for each sequence length,
we generate 10 random circuits.
The probability of measuring the expected survival state as a function
of sequence length is best fit to the model
\begin{equation}\label{eq: RB decay}
    p(L) = af^L+1/4,  
\end{equation}
and the average fidelity of the logical error channel per Clifford is then computed as
\begin{equation}\label{eq: avg fid}
    F_{avg}=1 - 3(1-f)/4.
\end{equation}
We divide the infidelity by the average number of CNOT gates per
Clifford group element in our construction to arrive at an estimate of the average fidelity per logical CNOT.
The decay plots are shown in Fig.~\ref{TQRB},
along with a comparison to the device emulators.
We obtain average fidelities per logical CNOT of 0.9991(2) and 0.9980(8) for H1-1 and H2-1, respectively.
For comparison, the physical native two-qubit gate
has been benchmarked at an average fidelity of 0.99912(3) on H1-1 and 0.99817(5) on H2-1,
when averaged over the gate zones~\cite{github_spec, Moses2023}.
This native gate is given by $U_{2Q}=e^{-iZZ\pi/4}$ and differs from CNOT only by single-qubit unitaries,
whose error is on the order of $\sim10^{-5}$ and negligible compared to the two-qubit gate error.
The logical CNOT fidelities are therefore competitive with the physical native two-qubit gates.
The higher logical fidelity on H1-1 compared to H2-1 
can be attributed to higher fidelity two-qubit gates,
as well as a larger memory error per depth-one circuit time on H2-1 due to the larger number of qubits in the QCCD architecture.
Here, memory error refers qubit
decoherence during the time required to transport ions~\cite{Moses2023}.

\subsection{Benchmarking of Logical \textit{T}}

We use the protocol in Ref.~\cite{Piveteau2021} to estimate
the error rate of the logical $T$ gate.
By inserting a twirl over the set $\{I,SX\}$ after the
injected $\ket{T}$ state in the circuit diagram of Fig.~\ref{fig: teleporation gadget2},
it was shown that the noise process on the logical
$T$ operator is
given by a dephasing channel
\begin{equation}\label{eq: T error channel}
\tilde \cT = (1-\epsilon)\cT + \epsilon\cZ\circ\cT,
\end{equation}
where $\cT$ and $\tilde \cT$ are the ideal and noisy channels,
and $\epsilon$ is the state infidelity of the injected state $\ket{T}$.

The parameter $\epsilon$ can be measured by preparing
$\ket{+}$, applying the noisy $T$ gate $L$ times
with $L$ a multiple of 4, and measuring in the logical $X$ basis.
The probability of obtaining the expected outcome is then fit to the model
\begin{equation}\label{eq: T decay}
    p(L) = \frac{1}{2} + \frac{1}{2}(1-2\epsilon)^L.
\end{equation}
The motivation in Ref.~\cite{Piveteau2021} for estimating
$\epsilon$ was that one can then apply error mitigation
via probabilistic error cancellation~\cite{Temme2017} on the logical $T$ gates.
We leave a demonstration of this technique to future work.

\begin{figure}[h]
\includegraphics[width=0.45\textwidth]{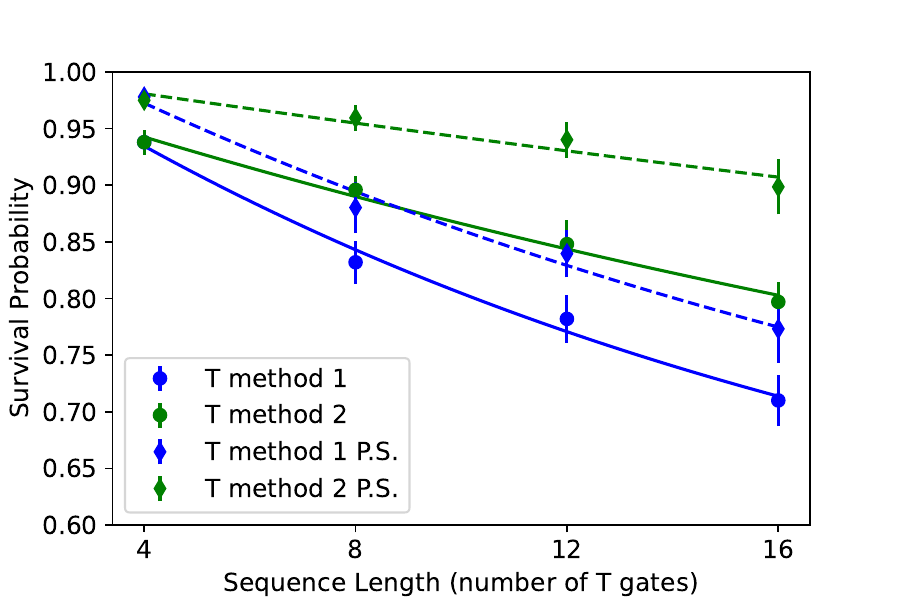}
\caption{
Logical $T$ gate benchmarking decay curves
for the device H2-1
and the two methods for the teleportation gadget
shown in Fig.~\ref{fig: teleporation gadget2}.
Solid lines are fit to the raw data and dashed lines to the data including post-selection on the
syndromes obtained from the logical measurement in the teleportation gadgets (labeled P.S.).
Ten random circuits per sequence length were chosen
and all circuits were run with 100 shots and submitted in a random order.
The curves are fit to Eq.~\eqref{eq: T decay},
and the estimated average fidelities per logical
$T$ are listed in Table~\ref{tab: benchmarks}.
}
\label{T_bench}
\end{figure}

We run the logical $T$ gate benchmarking experiment on the device H2-1 using both versions of the teleportation gadget shown in Fig.~\ref{fig: teleporation gadget2},
which we call methods one and two.
The circuits for the experiment use 15 physical qubits:
7 for each logical qubit block in Fig.~\ref{fig: teleporation gadget2},
plus one ancilla for fault-tolerant initialization of the data logical qubit.
The decay curves from the experiment are shown in Fig.~\ref{T_bench}.
We select $L\in\{4, 8, 12, 16\}$ and sample 10
random circuits per sequence length and run each
circuit for 100 shots.
After $\epsilon$ is determined from Eq.~\eqref{eq: T decay},
the average fidelity is computed as
\begin{equation}
    F_{avg} = 1-2\epsilon/3.
\end{equation}
We obtain estimated average fidelities per logical $T$ gate of 0.981(1) and 0.990(1)
for methods one and two, respectively.
We conjecture that method two performs better because
the logical state is swapped between the data qubit 
block and a fresh ancilla qubit block with each application of the gate.
This may lead to less accumulation of memory error on
the logical state.
This conjecture is supported by simulations from the device emulator,
which predicts a lower average fidelity for method one.
The emulator, in addition to modeling gate error,
contains detailed information about the ion transport schedule and models memory error incurred during transport.

If we post-select the data based on the syndrome information obtained from the logical measurement in the teleportation-gadget,
we obtain higher average fidelities: $0.985(3)$ and $0.995(1)$, respectively.
At the longest sequence length of $L=16$,
an average of 47\% of shots were retained over the two experiments.
However, as noted in Sec.~\ref{sec: 2},
this form of post-selection is not scalable,
but is still useful for near-term logical circuit experiments with a relatively small non-Clifford gate count.


\section{Logical QFT benchmarking}\label{sec: 4}
\begin{figure}[h]
\centering
\includegraphics[width=0.3\textwidth]{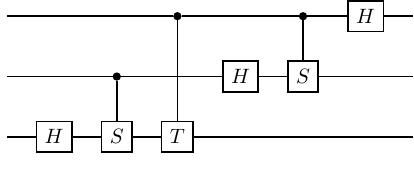}
\caption{
Three-qubit QFT circuit, excluding a final qubit reordering.
}
\label{fig: QFT}
\end{figure}

\begin{figure}[h]
\centering
\includegraphics[width=0.4\textwidth]{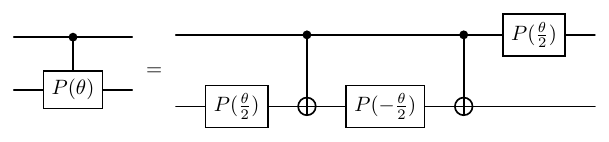}
\caption{
Control-phase gate decomposition.
Here $P(\frac{\pi}{2})=S$, and $P(\frac{\pi}{4})=T$.
}
\label{fig: control-phase}
\end{figure}

In this section, we describe our implementation of
the three-qubit QFT circuit, and present the results of our benchmark on the H2-1 quantum computer.
An overview of the H-series hardware is given in Appendix~\ref{appendix A}.
The three-qubit QFT circuit is shown in Fig.~\ref{fig: QFT}.
Note that we do not include the final swap of the qubits,
since this can be achieved by a relabeling of the qubits in software.
The circuit contains two control-$S$ gates,
and a control-$T$ gate.
We implement the control-$S$ gates using a standard decomposition into two CNOT gates and three $T$ or $T^{\dag}$ gates~\cite{Barenco1995},
which is shown in Fig.~\ref{fig: control-phase}.
We implement the control-$T$ in two different ways.
The first method makes use of a device that we call a recursive-teleportation gadget,
and the second uses a logical ancilla with a logical mid-circuit measurement and conditional $CZ$ gate.
In the following subsections we first describe these two implementations,
and then the protocol that we use to benchmark the QFT circuit before presenting the experimental results.
The total physical resources (qubits and two-qubit gates) for the two methods are listed in Table~\ref{tab: QFT resources}.

\begin{table}[h]
\begin{ruledtabular}
\begin{tabular}{lcccc}
$CT$ Method & Log. $\ket{\theta}$ & Log. TQ  & Phys. TQ & Phys. Qubits \\
\hline
Recursive & 9-12 & 15-18 & 237-291 & 28  \\
Ancilla & 11 & 13-14 & 256-263 & 28  \\
\end{tabular}
\end{ruledtabular}
\caption{Logical and physical resources required
for logical three-qubit QFT.
Here, log.~$\ket{\theta}$ indicates the number of 
logical non-Clifford state injections,
and TQ stands for two-qubit gates.
The ranges of values indicate the minimum and maximum 
possible number of gates due to the conditional gates.}
\label{tab: QFT resources}
\end{table}

\subsection{Control-\textit{T} by Recursive-Teleportation Gadget}
As shown in Fig.~\ref{fig: control-phase},
a control-$S$ gate can be decomposed into the Clifford+$T$ gate
set using a standard decomposition~\cite{Barenco1995}.
The control-$T$ gate is trickier. Using
the same decomposition, the control-$T$
requires three $P(\frac{\pi}{8})$ gates,
where $P(\theta)=\mathrm{diag}(1,e^{i\theta})$.
These are non-Clifford gates that perform a $Z$-rotation
by half as large an angle as the $T$ gate.
While the $P(\frac{\pi}{8})$ gate can be approximately decomposed into the Clifford+$T$ gate
set using synthesis algorithms~\cite{Ross2016},
such decompositions are quite costly in the number of $T$ gates required.
Our method for performing a logical $P(\frac{\pi}{8})$ gate
in the color code is to apply the same encoding circuit for the logical $\ket{T}$ state,
except with last physical qubit in Fig.~\ref{fig: encoding circ} prepared in the state $\ket{\pi/8}=P(\frac{\pi}{8})\ket{+}$.
We note that this preparation is non-fault-tolerant,
and unlike the logical $\ket{T}$ state preparation,
which can be made fault-tolerant by measuring the projector onto the logical state with a flag qubit~\cite{Chamberland2019},
we do not know how to make the $\ket{\pi/8}$ preparation fault-tolerant.
However, we note there are recent proposals for performing fault-tolerant logical $Z$-rotations by post-processing of results to interpolate to a desired rotation angle~\cite{Koczor2024},
or using quantum error detection along with post-selection for arbitrary-angle state injection~\cite{Choi2023}.

\begin{figure}[h]
\centering
\includegraphics[width=0.45\textwidth]{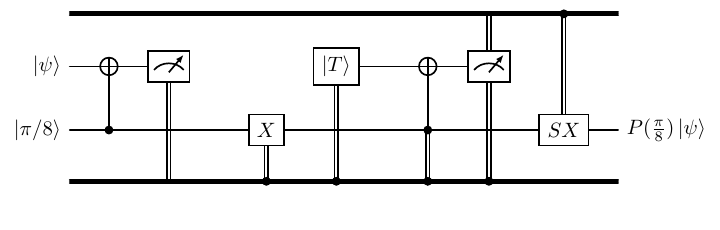}
\caption{
Recursive-teleportation gadget for performing a $P(\pi/8)$ gate.
The logical $\ket{\pi/8}$ state is prepared non-fault-tolerantly by the encoding circuit in Fig.~\ref{fig: encoding circ}.
After the first measurement, a teleportation gadget implementing the $T$ gate is performed
conditioned on the classical outcome.
Classical control wires are shown for clarity.
}
\label{fig: teleporation gadget3}
\end{figure}

After preparing the state $\ket{\pi/8}$,
a teleportation gadget as in Fig.~\ref{fig: teleporation gadget2} is applied to convert
the state into a $P(\frac{\pi}{8})$ gate. The teleportation gadget in this case has a conditional $T$ gate,
which itself must be performed by a teleportation gadget,
as shown in Fig.~\ref{fig: teleporation gadget3}.
This is why we call the technique the recursive-teleportation method.
This method can be scaled to higher levels of recursion,
where in general $P(\pi/2^k)$ is performed by
preparing a $\ket{\pi/2^k}$ state and using a teleportation gadget with a conditional $P(\pi/2^{k-1})$ gate.
In this way a general $n$-qubit QFT circuit can be encoded
logically if one is willing to sacrifice fault-tolerance.

\subsection{Ancilla-Assisted Control-\textit{T}}

\begin{figure}[h]
\centering
\includegraphics[width=0.5\textwidth]{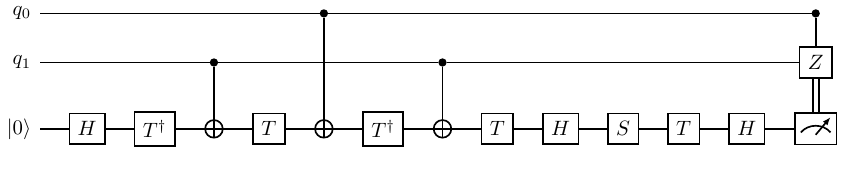}
\caption{
Ancilla-assisted control-$T$ circuit.
The circuit equals control-$T$ with control $q_0$ and target $q_1$.
}
\label{fig: ancilla assisted}
\end{figure}

The other method we use for implementing a logical control-$T$
makes use of a circuit construction from Ref.~\cite{Nam2020}, which is shown in Fig.~\ref{fig: ancilla assisted}.
The figure shows how with an ancilla qubit,
a controlled-$P(\theta)$ gate can be replaced by a $P(\theta)$,
along with a Clifford+$T$ circuit and a mid-circuit measurement and conditional $CZ$ gate.
The advantage of this construction for control-$T$ is that
it yields a Clifford+$T$ circuit, which can be implemented fault-tolerantly using methods of Ref.~\cite{Chamberland2019}.
The disadvantage is that for control-$P(\pi/2^k)$
gates with $k>2$, the construction no longer results
in a Clifford+$T$ circuit.

Naively, implementing the three-qubit QFT with the
ancilla-assisted control-$T$ requires 5 logical qubit blocks:
3 data qubits plus one ancilla for the control-$T$
plus another for the $T$ gate teleportation gadgets.
However, the QFT can be simulated with only 4 logical qubit blocks by first using one of the data qubits
as the ancilla for the control-$T$ on the other two,
before resetting that qubit as a data qubit for the remainder of the QFT circuit.
The full logical circuit therefore uses $4\times7=28$ physical qubits,
which fits within the 32-qubit architecture of H2-1.
We implement the ancilla-assisted control-$T$ gate 
using method two for the $T$ gate teleportation gadget shown in Fig.~\ref{fig: teleporation gadget2},
since this method achieved the higher fidelity.
We benchmark this logical control-$T$ by applying
it to each basis state from two bases
(see Appendix~\ref{appendix D}),
obtaining an average output state fidelity of $0.79(2)$.

\subsection{Fidelity Lower Bounds for QFT}
We benchmark the logical QFT by applying it to each basis state in 
a pair of mutually unbiased bases,
which is sufficient to obtain a lower bound on the process fidelity~\cite{Hofmann2005}. A general explanation of this procedure is as follows.

Let $U$ be the $n$-qubit QFT, which is defined by
\begin{equation}
    U\ket{x}=\frac{1}{\sqrt{d}}\sum_{y=0}^{d-1}e^{2\pi i xy/d}\ket{y}.
\end{equation}
A short calculation shows that the state $U\ket{x}$ is separable:
\begin{equation}
    U\ket{x}=\bigotimes_{j=0}^{n-1}\frac{1}{\sqrt{2}}\big(\ket{0}+e^{2\pi i x/2^{n-j}}\ket{1}\big).
\end{equation}
Therefore, the
fidelity of a state with respect to $U\ket{x}$
can be measured using only single-qubit gates.
Specifically, applying $HR_z(-2\pi x/2^{n-j})$ to the $j$-th 
qubit maps the ideal state back to the all-zero state.
Next, define the Fourier basis to be $\{\ket{f_x}=U\ket{x}\}_{x=0}^{d-1}$.
The computational basis and Fourier basis form a pair of mutually unbiased bases~\cite{Wooters1989}, since for all $x,y\in\{0,\dots,d-1\}$, we have
\begin{equation}
    \abs{\braket{y}{f_x}}=\frac{1}{\sqrt{d}}.
\end{equation}
In the same way that the outputs of the QFT on the computational
basis states can be measured using only single-qubit gates,
the Fourier basis states can be prepared using only single-qubit gates.
It can be shown that the QFT acting on a Fourier basis state
gives a computational basis state, specifically
\begin{equation}
    U\ket{f_x}=\ket{-x\,\,(\mathrm{mod}\,\,d)}.
\end{equation}
It follows that the output state fidelity after applying the QFT to a Fourier basis state can be measured with a simple computational basis measurement.
We can now make use of a theorem due to Hofmann,
which gives a lower bound on the process fidelity of a process $\cE$ in terms of the average of its output state fidelities,
averaged over two mutually unbiased bases~\cite{Hofmann2005}.
The result is that if $F_1$ and $F_2$ are the averages of the output
state fidelities in the two bases, that is,
\begin{align}
F_1&=\frac{1}{d}\sum_{x=0}^{d-1}\bra{x}U^{\dag}\cE(\ket{x}\bra{x})U\ket{x},\notag\\
F_2&=\frac{1}{d}\sum_{x=0}^{d-1}\bra{f_x}U^{\dag}\cE(\ket{f_x}\bra{f_x})U\ket{f_x},
\end{align}
then the process fidelity of $\cE$
is lower bounded according to
\begin{equation}\label{eq: fid bounds}
    F_{pro}(\cE, U)\ge F_{lo} = F_1+F_2-1.
\end{equation}
We therefore benchmark the QFT by preparing the $d$ computational 
basis states, and the $d$ Fourier basis states, applying the QFT to
each state and compute the lower bound in Eq.~\eqref{eq: fid bounds}.
Note that the bound is stated in terms of the process fidelity.
The average fidelity, which is how we reported the      
logical component fidelities in Sec.~\ref{sec: 3},
is related to the process fidelity by
\begin{equation}
    F_{avg}=(dF_{pro}+1)/(d+1),
\end{equation}
where $d=8$ for the three-qubit QFT~\cite{Nielsen2002}.

Although all state preparations and measurement settings in this protocol
require only single-qubit gates,
for exactly half of the input states the measurement setting in the 
computational basis or the state preparation in the Fourier basis requires a $T$ or $T^{\dag}$ gate.
Therefore, for these input states the fidelity measurement
will include a logical SPAM error roughly equal to the logical ${T}$ gate error.
However, this error is still small compared to
the overall circuit error, as the circuit contains between 9 and 12 $T$ gates, depending on the method.
The results that we report in the next section do not correct for this SPAM error.

Finally, we note that while Eq.~\eqref{eq: fid bounds} gives a
rigorous lower bound on the process fidelity,
this bound is often quite loose in practice. 
For example, a process with a global 
depolarizing error will have an average fidelity $F_{avg}=F_1=F_2$.
While we do not expect our QFT circuits to have a global depolarizing error,
in our discussion of the results we will focus more on the average output state fidelities $F_1$ and $F_2$ than on the lower bound.
However, we stress that it is important to apply the circuit to both bases in order
to obtain a lower bound,
since there are processes that act perfectly on one of the bases but still have a process fidelity of zero with respect to the target.

\subsection{Experimental Results}

\begin{figure*}
        \centering
        \subfloat[]{
            \centering
            \includegraphics[width=0.475\textwidth]{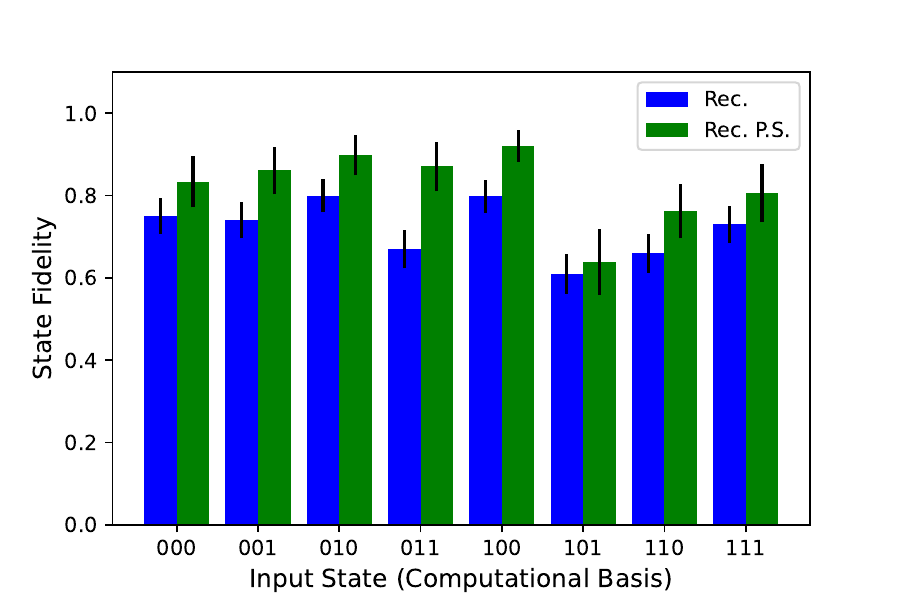}}
        \hfill
        \subfloat[]{
            \centering 
        \includegraphics[width=0.475\textwidth]{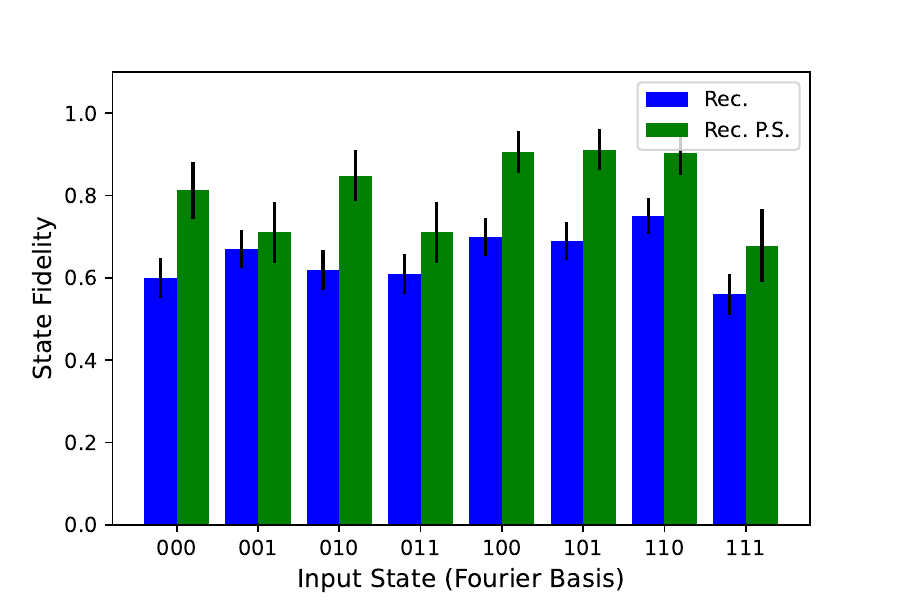}}
        \vfill
        \subfloat[]{
            \centering
            \includegraphics[width=0.475\textwidth]{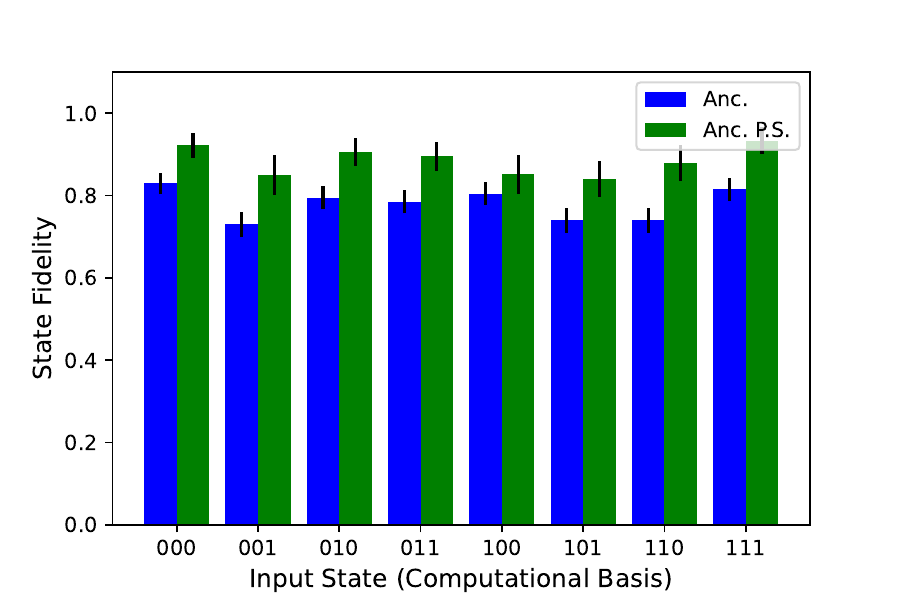}}
        \hfill
        \subfloat[]{
            \centering 
            \includegraphics[width=0.475\textwidth]{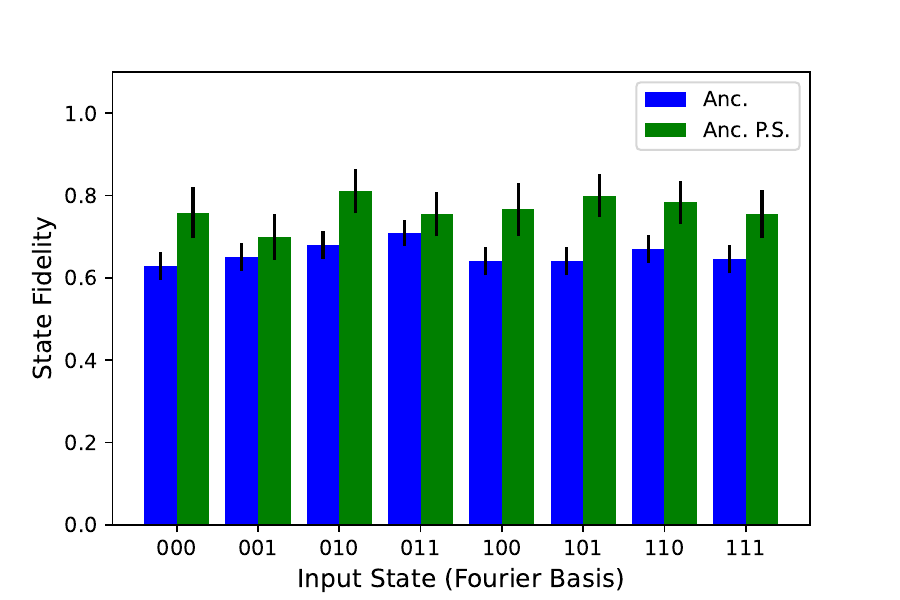}}
        
        \caption[]
        {Logical QFT output state fidelities on the H2-1 device for
        (a) recursive-teleportation method, computational basis, (b) recursive-teleportation method, Fourier basis,
        (c) ancilla-assisted method, computational basis,
        (d) ancilla-assisted method, Fourier basis.
        The blue and green bars are without and with
        post-selection on the syndromes from the measurements in the $T$ gate teleportation gadgets,
        as explained in Sec.~\ref{sec: 2}.
        The circuits were run with 100 and 200 total shots for the recursive-teleportation and ancilla-assisted methods, respectively.
        }
        \label{fig: QFT bar plots}
    \end{figure*}

We benchmark the three-qubit logical QFT circuits on the H2-1 quantum computer~\cite{Moses2023},
using both the recursive-teleportation and the ancilla-assisted control-$T$ methods.
We apply the logical QFT to each basis state
in the computational and Fourier bases,
and run each logical circuit for 100 shots.
For the ancilla-assisted method, we average two full data sets taken roughly one week apart.
Bar plots of the state fidelities for each circuit are shown in Fig.~\ref{fig: QFT bar plots},
and the quantities $F_1$ and $F_2$ as well as the
average fidelity lower bounds are summarized in Table~\ref{QFT_table}.
We find that the computational basis outperforms the Fourier basis for both methods,
with $F_1=0.72(2)$ and $F_2=0.65(2)$ for the recursive-teleportation method and $F_1=0.78(1)$ and $F_2=0.66(1)$
for the ancilla-assisted method.
If we post-select on the measurements in the teleportation gadgets,
the fidelities increase, up to $0.89(1)$ for the computational basis with the ancilla-assisted method.
The fraction of shots retained when post-selecting were 0.35(1) for the the recursive-teleportation method and 0.31(1) for the ancilla-assisted method.
Interestingly, this post-selection removes the bias between the computational and Fourier bases for the recursive-teleportation method,
but not the ancilla-assisted method.
We conclude that quantum error detection of this form,
while not scalable,
can be useful for improving the performance of near-term logical circuit experiments.

\begin{table}[h]
\begin{ruledtabular}
\begin{tabular}{lccc}
Experiment & $F_1$ & $F_2$ & $F_{avg}$ bound\\
\hline
Recursive-teleportation & 0.72(2) & 0.65(2) & 0.44(2)    \\
Recursive-teleportation (P.S.) & 0.82(2) & 0.81(2) & 0.68(3)    \\
Ancilla-assisted  & 0.78(1) & 0.66(1) & 0.50(1) \\
Ancilla-assisted (P.S.) &  0.89(1) & 0.77(2) & 0.69(2) \\
\end{tabular}
\end{ruledtabular}
\caption{QFT output state fidelities.
$F_1$ and $F_2$ are the average output state fidelity
for the computational and Fourier bases, respectively.
Post-selected (P.S.) refers to discarding shots where the 
measurement in the teleportation gadget detected an error,
as explained in Sec.~\ref{sec: 2}.}
\label{QFT_table}
\end{table}


Given the complexity of the QFT circuits run
in this experiment,
it is useful to ask whether the results can be accounted for by the logical component errors.
From Table.~\ref{tab: benchmarks},
we see that the logical $T$ gate is the leading contributor to circuit error with an average fidelity 0.990(1),
compared to 0.9980(8) for logical CNOT.
We can obtain a heuristic estimate for an output state fidelity of the QFT circuit
by assuming single-qubit and two-qubit depolarizing error channels for the logical $T$ and two-qubit gates.
A depolarizing channel with parameter 
$\epsilon$ acts on a state $\rho$ according to
\begin{equation}
    \cD(\rho)=(1-\epsilon)\rho+\epsilon I/d,
\end{equation}
where the depolarizing parameter is related to the average fidelity by
\begin{equation}
    1-F_{avg} = \frac{d-1}{d}\epsilon.
\end{equation}
We use Qiskit~\cite{Qiskit} to compute the
output state fidelities of the ancilla-assisted QFT circuit, applied to the computational and Fourier bases, and 
assuming a depolarizing error model for the logical components.
We only model the QFT via the ancilla-assisted method,
since this yields a Clifford+$T$ circuit,
whereas the recursive-teleportation gadget implements a logical $P(\frac{\pi}{8})$ gate that we did not benchmark,
though we expect the logical infidelity of the $\ket{\pi/8}$ state preparation to be comparable to that of the $\ket{T}$ state.
From our model, we obtain average output state fidelities of $F_1=0.898$ and $F_2=0.889$. 
The experimental values for $F_1$ and $F_2$ are
noticeably lower than the heuristic prediction,
indicating additional error that is not accounted for by the component level benchmarks alone.
Our hypothesis is that memory error during ion transport
leads to system-level error that is not present in the component
benchmarking experiments.


\section{Conclusion}\label{sec: 5}
In this work, we have reported the first demonstration of a logical three-qubit QFT circuit,
as part of our
efforts at characterizing the performance of circuits encoded in a QEC code.
Learning from the development of quantum processors at the physical level, the path toward high performance at the logical level will likely require the careful study of the different components afforded by different encodings, such as the gate sets, syndrome extraction procedures,
and magic state injection methods.

While this study is not comprehensive for even the particular code used in this investigation,
our results show that the choices between different implementations of
logical components matter.
For example, we found that on the H2-1 device, the $T$ gadget teleportation method two
from Fig.~\ref{fig: teleporation gadget2} was the superior one.
Overall, the non-Clifford gate operations are the dominant source of error.
We note that an accounting of the error sources using a gate level error model along with their respective measured component error rates does not match the total error observed in the system-level benchmark of the QFT. These estimates would predict state fidelities just below $0.9$,  whereas the measurements were at most $0.78$.
This discrepancy could be due to several omissions, including memory error, leakage errors, or the coherent accumulation of noise. 

 
These results suggest at least two future areas of study: (1) it is imperative that we close the gap in our understanding between the component-level and system-level error rates on the logical level, and (2) we must find better methods for implementing non-Clifford operations through a combination of hardware improvements and code developments including higher code distances and 
the eventual use of fully fault-tolerant operations.
Experimental fault-tolerant quantum computing still has many challenges ahead. There are different codes to explore, different protocols within those codes, and all of them will have different strengths and weaknesses in the context of real hardware that can be built.
Sorting through these methods is the high-level challenge, but before that, researchers need to develop the tools to make that sorting feasible.

We hope this work, with the development of logical benchmarking enabled by high-level programming tools, lays a good foundation for the experimental side of this challenge.
Furthermore, we hope that this work opens the door to implementing other algorithms based on the QFT at the logical level,
such as solving the one-dimensional wave equation~\cite{wright2024}.
Finally, we believe tracking the state-of-the-art performance of a logical QFT in real quantum processors would make a valuable system-level benchmark for logical encodings in quantum processors, one that we hope the larger community can adopt.

\section*{Code Availability}
The OpenQasm programs that were submitted to the H-series hardware,
and the code for analyzing the data and making the plots in this paper,
are available here~\cite{github_repo}.
The SLR code base, which was used to create the logical circuits and convert the circuits to OpenQasm,
is available here~\cite{PECOS}. 

\acknowledgements{
We thank Ben Criger for early discussions and suggestions.
We thank the entire Quantinuum hardware team for
making these experiments possible.
The ion traps used in this work were fabricated by Honeywell.
}

\appendix

\section{Hardware Overview}\label{appendix A}

The experiments were run on the Quantinuum H1-1 and H2-1 devices. H1-1 is a 20-qubit trapped-ion QCCD quantum computer that operates on a linear surface trap described in Ref.~\cite{Pino2020}. H2-1 is a 32-qubit trapped-ion QCCD quantum computer that operates on a racetrack surface trap described in Ref.~\cite{Moses2023}.
Both systems use ground state hyperfine levels of $^{171}\textrm{Yb}+$ ions
as physical qubits,
along with $^{138}\textrm{Ba}+$ ions for sympathetic cooling.
The systems feature physical native single-qubit (1Q) and two-qubit (2Q) gates given
by $U_{1Q}(\theta, \phi)=\exp[-i\theta(X \cos \phi + Y \sin \phi)/2]$, and $U_{2Q}(\theta)=\exp[-i\theta ZZ/2]$,
and allow for mid-circuit measurement and reset with low crosstalk errors~\cite{Gaebler2021}.
Ion transport, which includes linear shifts, physical swaps, and split combines, allows for arbitrary rearrangement and pairing but induces small memory errors (relative to the 2Q gate error). Table~\ref{tab:hardware} contains the relevant error magnitudes for different component operations. The tests used to measure each error magnitude are further described in Ref.~\cite{Moses2023}. Complete datasets are given in Ref.~\cite{github_spec}.

\begin{table}[] 
\begin{ruledtabular}
\begin{tabular}{lcc}
Error           &  H1-1        & H2-1   \\ \hline
Single-qubit              &  0.29(5)     & 0.25(3)        \\
Two-qubit              &  8.8(3)     & 18.3(5)             \\       
Memory          & 1.8(2)       & 2.2(3)           \\
Measurement crosstalk & 0.083(9)  & 0.045(6)      \\
State preparation and measurement  & 26(1)  & 16(1)          \\
\end{tabular}
\end{ruledtabular}
\caption{Error parameters for H1-1 and H2-1 devices used for experiments and all values are $\times 10^{-4}$.
Single-qubit and two-qubit gate error is defined as one minus average fidelity.
}
\label{tab:hardware}
\end{table}

\section{Simulation of Partially-Fault-Tolerant Logical \textit{T} Benchmarking}\label{appendix B}

\begin{figure}[h]
\centering
\includegraphics[width=0.35\textwidth]{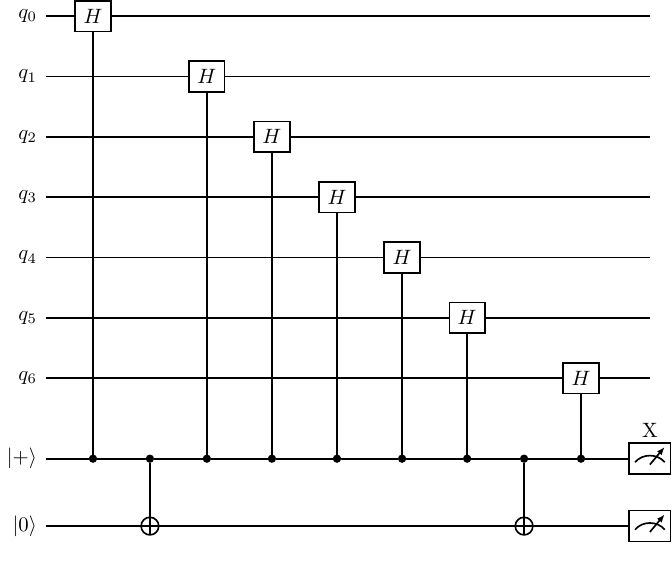}
\caption{
Logical Hadamard measurement with a flag qubit,
used to make logical state preparation fault-tolerant when followed by a round of QEC.
The measurement of the first ancilla records the value of the logical Hadamard,
and the second ancilla is a flag qubit, for which a result of `1' indicates an
error that can spread to an uncorrectable error.
The encoding circuit in Fig.~\ref{fig: encoding circ} along with the
circuit above is repeated until both ancilla qubits are measured as `0'.
}
\label{fig: Hadamard measurement}
\end{figure}

To prepare the state $\ket{\overline {T}}$,
we first prepare the state
\begin{equation}
    \ket{\overline {H}}=\cos(\pi/8)\ket{\overline {0}} + \sin(\pi/8)\ket{\overline {1}},
\end{equation}
from which $\ket{\overline {T}}$ is obtained via logical Clifford gates, according to
\begin{equation}
    \ket{\overline {T}} =\overline{R_z(\pi/2)}\, \overline{R_x(\pi/2)}\ket{\overline {H}}.
\end{equation}
As shown in Ref.~\cite{Chamberland2019},
the state $\ket{\overline{H}}$ can be prepared fault-tolerantly
by first using the circuit in Fig.~\ref{fig: encoding circ} to prepare $\ket{\overline {H}}$ non-fault-tolerantly,
and then measuring $\overline{H}$
with the use of a flag qubit to detect bad errors,
and finally performing a round of QEC.
If the final QEC round is omitted,
but the flagged $\overline{H}$ measurement is still made,
the procedure is called partially-fault-tolerant
and was demonstrated in Ref.~\cite{Postler2022}.
Since $\overline{H}^2=I$ and $\overline H$ is Clifford,
$\overline{H}$ can be measured by applying a control-$H$ between an ancilla physical qubit prepared
in $\ket{+}$ and each physical qubit in the logical code block,
and then measuring the ancilla in the $X$ basis.
This circuit is shown in Fig.~\ref{fig: Hadamard measurement}.
If the ancilla qubit and the flag qubit are both measured as `0',
the fault-tolerant preparation succeeds and $\ket{\overline{T}}$ is input into the teleportation 
gadget in Fig.~\ref{fig: teleporation gadget2} to implement a $\overline{T}$ gate.
The $\ket{\overline{H}}$ state preparation procedure can be
performed as a repeat-until-success protocol,
with a limit on the number of repetitions.
If the limit is reached without a successful
state preparation, then that shot is discarded in post-selection,
though we note that in a larger quantum computer,
a magic state factory could prepare multiple states in
parallel so that there is always a resource state ready
when a $\overline{T}$ gate is to be performed.

\begin{figure}[h]
    \centering
    \includegraphics[width=0.45\textwidth]{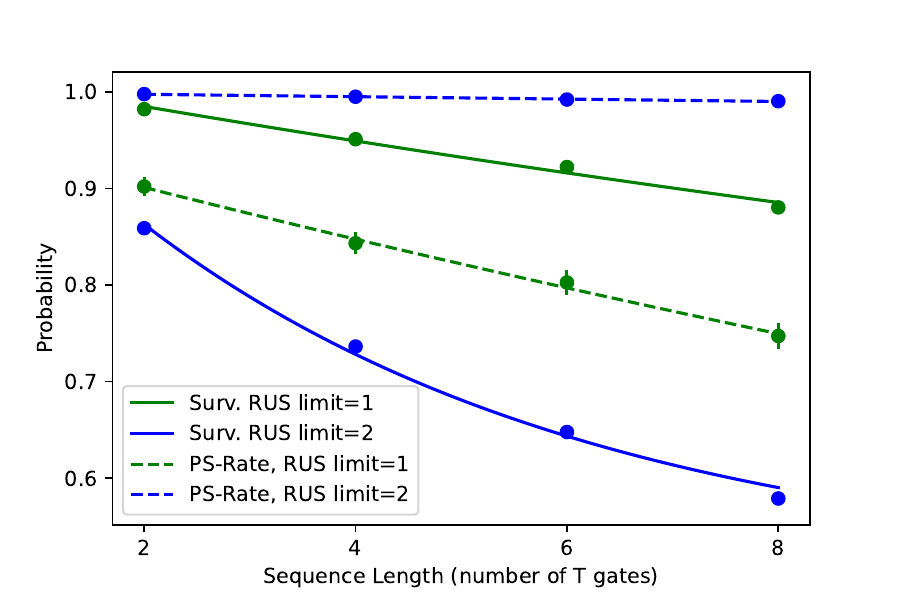}    
        \caption[]
        {Simulation of partially fault-tolerant logical $T$ gate benchmarking on H1-1E.
        The fault-tolerant gate uses a repeat-until-success (RUS)
        protocol,
        with a limit to the number of repetitions,
        chosen from $\{1,2\}$.
        The solid and dashed lines are
        survival probability and post-selection rate, respectively.
         10 random circuits were chosen per sequence length.
        Each circuit was run for 1000 shots.} 
        \label{fig: FT T sim}
\end{figure}

We benchmark the partially-fault-tolerant $\overline{T}$ gate on the
device emulator H1-1E, using the teleportation gadget method two in Fig.~\ref{fig: teleporation gadget2},
and the same benchmarking procedure as in Sec.~\ref{sec: 3} for the non-fault-tolerant $\overline{T}$,
with the RUS limit chosen from $\{1,2\}$.
The decay curves and post-selection rate,
which is the fraction of shots retained
are shown in Fig.~\ref{fig: FT T sim}.
The plot shows that RUS limit one achieves a higher average fidelity,
since there are fewer gates on average,
at the cost of more post-selection needed.
The average fidelities and post-selection rates are listed in Table.~\ref{tab: FT T}.
Given the rate of post-selection,
and the fact that the average fidelity per logical $T$ was simulated to be slightly lower than our 
experimentally measured non-fault-tolerant fidelity,
we chose to implement the QFT circuits using a non-fault-tolerant logical $T$.

\begin{table}[h]
\begin{ruledtabular}
\begin{tabular}{lcc}
RUS limit & Average Fidelity & Post-selection Rate ($L=8$) \\
\hline
1 & 0.988(1) &  0.75(1) \\
2 & 0.931(4) &  0.990(3) \\
\end{tabular}
\end{ruledtabular}
\caption{Average fidelity of logical $T$ and post-selection rate (fraction of shots retained) at sequence length 8, 
for different repeat-until-success (RUS) limits,
from simulation on the emulator H1-1E.
The fidelity is obtained from best-fitting the decay curves in Fig.~\ref{fig: FT T sim}
}
\label{tab: FT T}
\end{table}

\section{SLR Example}\label{appendix C}

In this section we walk through an SLR program
that prepares a logical qubit in the logical $\ket{+}$ state,
applies logical $T$ via a teleportation gadget,
and finally measures in the logical $X$ basis.
The first step is to initialize the SLR program with the
quantum and classical registers required for this task.

\begin{lstlisting}[language=python, basicstyle=\small]
program = SLR(

    q := QReg('q', 7), # logical qubit
    a := QReg('a', 7), # logical ancilla

    # classical registers for output
    c := CReg('c', 7),
    cLogRaw := CReg('cLogRaw', 1),
    cLog := CReg('cLog', 1),
    cSyn := CReg('cSyn', 3),

    # classical register for RUS init
    init := CReg('init', 1),
    
    # classical registers for T gadget
    cT := CReg('cT', 7),
    cTLogRaw := CReg('cTLogRaw', 1),
    cTLog := CReg('cTLog', 1),
    cTSyn := CReg('cTSyn', 3),
)
\end{lstlisting}
We then initialize the logical qubit using the fault-tolerant repeat-until-success protocol described in Sec.~\ref{sec: 2}.

\begin{lstlisting}[language=python, basicstyle=\small] 
# fault tolerant state initialization
program.extend(
    InitRUS(q, a[0], init[0], limit=3),
    H(q),
)
\end{lstlisting}
Here we have used the physical qubit \lstinline{a[0]} as the ancilla qubit
for the initialization.
The ancilla register will be reused for the $\ket{\overline{T}}$ preparation.
The RUS protocol succeeds if the classical
bit \lstinline{init[0]} has a value of `0'.
We then apply the $\overline{T}$ gate teleportation gadget.

\begin{lstlisting}[language=python, basicstyle=\small]

# T gate gadget
program.extend(  
    InitTPlusNonFT(a),
    CX(q,a),
    MeasDecode(a,cT,cTLogRaw,cTLog,cTSyn),
    If(cTLog[0]==1).Then(S(q)),
)

\end{lstlisting}
Here the \lstinline{MeasDecode} object represents a measurement of the logical qubit \lstinline{a}
into the classical 7-bit register \lstinline{cT},
with a 1-bit register \lstinline{cTLog} that contains
the logical binary value computed from \lstinline{cT},
including the syndrome information and decoding.
Finally, we measure the logical qubit \lstinline{q} in the $X$ basis.

\begin{lstlisting}[language=python, basicstyle=\small]

# measure
program.extend(
    H(q),
    MeasDecode(q,c,cLogRaw,cLog,cSyn),
)

\end{lstlisting}

\section{Fidelity Lower Bounds for Control-\textit{T}}\label{appendix D}

\begin{figure}[h]
    \centering
    \includegraphics[width=0.45\textwidth]{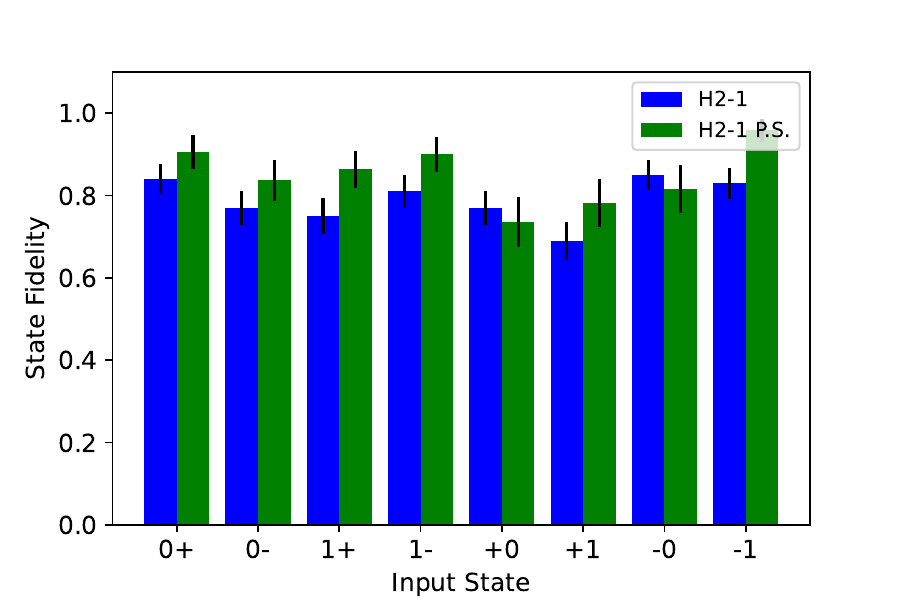}    
        \caption[]
        {Output state fidelities for logical control-$T$ on the device H2-1, via the
        ancilla-assisted method described in Sec.~\ref{sec: 4}.
        The blue (green) bars are without (with) post-selection on the syndromes from the measurements in the $T$ gate teleportation gadgets.
        All circuits were run for 100 shots.
        }
        \label{fig: control-T}
\end{figure}

The control-$T$ gate has the following action on the two-qubit bases of mixed $X$ and $Z$ eigenstates:
\begin{align}\label{eq: CT truth table 1}
    \ket{+}\ket{0} &\mapsto \ket{+}\ket{0}\notag\\
    \ket{+}\ket{1} &\mapsto \ket{T}\ket{1}\notag\\
    \ket{-}\ket{0} &\mapsto \ket{-}\ket{0}\notag\\
    \ket{-}\ket{1} &\mapsto \ket{T-}\ket{1}
\end{align}

\begin{align}\label{eq: CT truth table 2}
    \ket{0}\ket{+} &\mapsto \ket{0}\ket{+}\notag\\
    \ket{0}\ket{-} &\mapsto \ket{0}\ket{-}\notag\\
    \ket{1}\ket{+} &\mapsto \ket{1}\ket{T}\notag\\
    \ket{1}\ket{-} &\mapsto \ket{1}\ket{T-}
\end{align}
Here, the left-most ket is the state of the control qubit and $\ket{T-}=T\ket{-}$. These two bases form
a pair of mutually unbiased bases, so the process fidelity lower bound of Eq.~\eqref{eq: fid bounds} applies.
We implement the logical control-$T$ on H2-1 using the ancilla-assisted method described in Sec.~\ref{sec: 4}, and teleportation method two
for the $T$ gates.
We apply logical control-$T$ to each logical basis state in the two bases of mixed $X$ and $Z$ eigenstates.
As seen in Eq.~\eqref{eq: CT truth table 1} and Eq.~\eqref{eq: CT truth table 2},
for half of the input states the ideal output state includes a $\ket{T}$ or $\ket{T-}$.
In this case we apply a $T^{\dag}$ in order to measure the state fidelity,
and therefore the fidelity measurement has a SPAM error that is at least as large as that of a logical $T$ gate.
The output state fidelities are plotted in Fig.~\ref{fig: control-T},
where we do not correct for the logical SPAM error.
We obtain average state fidelities of $F_1=0.79(2)$ and
$F_2=0.79(2)$ over the two bases, from which we obtain a fidelity lower bound of $F_{avg}\ge0.66(2)$.
For comparison,
when we post-select on the syndrome information
from the measurement in the teleportation gadgets,
we obtain $F_1=0.82(3)$ and $F_2=0.88(2)$
with a lower bound $F_{avg}\ge0.76(3)$.

\nocite{*}

\bibliography{library}

\end{document}